\documentclass[conference]{IEEEtran}

\usepackage[acronyms,nonumberlist,nopostdot,nomain,nogroupskip,acronymlists={hidden}]{glossaries}
\usepackage{color}
\usepackage{amsmath}
\usepackage{subcaption}

\usepackage{graphicx}
\usepackage{multirow}
\usepackage{tabularx}
\usepackage{comment}
\usepackage{soul}
\usepackage{float}
\usepackage{balance}
\usepackage{url}

\ifCLASSINFOpdf
\else
\fi
\begin{document}
\newacronym{3gpp}{3GPP}{3rd Generation Partnership Project}
\newacronym{4g}{4G}{4th generation}
\newacronym{5g}{5G}{5th generation}
\newacronym{6g}{6G}{6th generation}
\newacronym{5gc}{5GC}{5G Core}
\newacronym{aau}{AAU}{Active Antenna Unit}
\newacronym{abf}{ABF}{Analog Beamforming}
\newacronym{dbf}{DBF}{Digital Beamforming}
\newacronym{adc}{ADC}{Analog to Digital Converter}
\newacronym{aerpaw}{AERPAW}{Aerial Experimentation and Research Platform for Advanced Wireless}
\newacronym{ai}{AI}{Artificial Intelligence}
\newacronym{aimd}{AIMD}{Additive Increase Multiplicative Decrease}
\newacronym{am}{AM}{Acknowledged Mode}
\newacronym{amc}{AMC}{Adaptive Modulation and Coding}
\newacronym{amf}{AMF}{Access and Mobility Management Function}
\newacronym{aops}{AOPS}{Adaptive Order Prediction Scheduling}
\newacronym{api}{API}{Application Programming Interface}
\newacronym{apn}{APN}{Access Point Name}
\newacronym{ap}{AP}{Application Protocol}
\newacronym{aqm}{AQM}{Active Queue Management}
\newacronym{ausf}{AUSF}{Authentication Server Function}
\newacronym{avc}{AVC}{Advanced Video Coding}
\newacronym{awgn}{AGWN}{Additive White Gaussian Noise}
\newacronym{balia}{BALIA}{Balanced Link Adaptation Algorithm}
\newacronym{bbu}{BBU}{Base Band Unit}
\newacronym{bdp}{BDP}{Bandwidth-Delay Product}
\newacronym{ber}{BER}{Bit Error Rate}
\newacronym{bf}{BF}{Beamforming}
\newacronym{bler}{BLER}{Block Error Rate}
\newacronym{brr}{BRR}{Bayesian Ridge Regressor}
\newacronym{bs}{BS}{Base Station}
\newacronym{bsr}{BSR}{Buffer Status Report}
\newacronym{bss}{BSS}{Business Support System}
\newacronym{ca}{CA}{Carrier Aggregation}
\newacronym{caas}{CaaS}{Connectivity-as-a-Service}
\newacronym{cav}{CAV}{Connected and Autonoums Vehicle}
\newacronym{cb}{CB}{Code Block}
\newacronym{cc}{CC}{Congestion Control}
\newacronym{ccid}{CCID}{Congestion Control ID}
\newacronym{cco}{CC}{Carrier Component}
\newacronym{cd}{CD}{Continuous Delivery}
\newacronym{cdd}{CDD}{Cyclic Delay Diversity}
\newacronym{cdf}{CDF}{Cumulative Distribution Function}
\newacronym{cdn}{CDN}{Content Distribution Network}
\newacronym{cli}{CLI}{Command-line Interface}
\newacronym{cn}{CN}{Core Network}
\newacronym{codel}{CoDel}{Controlled Delay Management}
\newacronym{comac}{COMAC}{Converged Multi-Access and Core}
\newacronym{cord}{CORD}{Central Office Re-architected as a Datacenter}
\newacronym{cornet}{CORNET}{COgnitive Radio NETwork}
\newacronym{cosmos}{COSMOS}{Cloud Enhanced Open Software Defined Mobile Wireless Testbed for City-Scale Deployment}
\newacronym{cots}{COTS}{Commercial Off-the-Shelf}
\newacronym{cp}{CP}{Control Plane}
\newacronym{cpe}{CPE}{Customer Premises Equipment}
\newacronym{cyp}{CP}{Cyclic Prefix}
\newacronym{up}{UP}{User Plane}
\newacronym{cpu}{CPU}{Central Processing Unit}
\newacronym{cqi}{CQI}{Channel Quality Information}
% \newacronym{cr}{CR}{Cognitive Radio}
\newacronym{cr}{CR}{Cell Reconfiguration}
\newacronym{cran}{C-RAN}{Centralized RAN}
\newacronym{crs}{CRS}{Cell Reference Signal}
\newacronym{csi}{CSI}{Channel State Information}
\newacronym{csirs}{CSI-RS}{Channel State Information - Reference Signal}
\newacronym{cu}{CU}{Central Unit}
\newacronym{d2tcp}{D$^2$TCP}{Deadline-aware Data center TCP}
\newacronym{d3}{D$^3$}{Deadline-Driven Delivery}
\newacronym{dac}{DAC}{Digital to Analog Converter}
\newacronym{dag}{DAG}{Directed Acyclic Graph}
\newacronym{das}{DAS}{Distributed Antenna System}
\newacronym{dash}{DASH}{Dynamic Adaptive Streaming over HTTP}
\newacronym{dc}{DC}{Dual Connectivity}
\newacronym{dccp}{DCCP}{Datagram Congestion Control Protocol}
\newacronym{dce}{DCE}{Direct Code Execution}
\newacronym{dci}{DCI}{Downlink Control Information}
\newacronym{dctcp}{DCTCP}{Data Center TCP}
\newacronym{dl}{DL}{Downlink}
\newacronym{dmr}{DMR}{Deadline Miss Ratio}
\newacronym{dmrs}{DMRS}{DeModulation Reference Signal}
\newacronym{drlcc}{DRL-CC}{Deep Reinforcement Learning Congestion Control}
\newacronym{dsrc}{DSRC}
{dedicated short-range communications}
\newacronym{d2d}{D2D}{device-to-device}
\newacronym{drs}{DRS}{Discovery Reference Signal}
\newacronym{du}{DU}{Distributed Unit}
\newacronym{e2e}{E2E}{end-to-end}
\newacronym{earfcn}{EARFCN}{E-UTRA Absolute Radio Frequency Channel Number}
\newacronym{ecaas}{ECaaS}{Edge-Cloud-as-a-Service}
\newacronym{ecn}{ECN}{Explicit Congestion Notification}
\newacronym{edf}{EDF}{Earliest Deadline First}
\newacronym{embb}{eMBB}{Enhanced Mobile Broadband}
\newacronym{empower}{EMPOWER}{EMpowering transatlantic PlatfOrms for advanced WirEless Research}
\newacronym{enb}{eNB}{evolved Node Base}
\newacronym{endc}{EN-DC}{E-UTRAN-\gls{nr} \gls{dc}}
\newacronym{epc}{EPC}{Evolved Packet Core}
\newacronym{eps}{EPS}{Evolved Packet System}
\newacronym{es}{ES}{Edge Server}
\newacronym{etsi}{ETSI}{European Telecommunications Standards Institute}
\newacronym[firstplural=Estimated Times of Arrival (ETAs)]{eta}{ETA}{Estimated Time of Arrival}
\newacronym{eutran}{E-UTRAN}{Evolved Universal Terrestrial Access Network}
\newacronym{faas}{FaaS}{Function-as-a-Service}
\newacronym{fapi}{FAPI}{Functional Application Platform Interface}
\newacronym{fdd}{FDD}{Frequency Division Duplexing}
\newacronym{fdm}{FDM}{Frequency Division Multiplexing}
\newacronym{fdma}{FDMA}{Frequency Division Multiple Access}
\newacronym{fed4fire}{FED4FIRE+}{Federation 4 Future Internet Research and Experimentation Plus}
\newacronym{fir}{FIR}{Finite Impulse Response}
\newacronym{fit}{FIT}{Future \acrlong{iot}}
\newacronym{fpga}{FPGA}{Field Programmable Gate Array}
\newacronym{fr2}{FR2}{Frequency Range 2}
\newacronym{fr1}{FR1}{Frequency Range 1}
\newacronym{fs}{FS}{Fast Switching}
\newacronym{fscc}{FSCC}{Flow Sharing Congestion Control}
\newacronym{ftp}{FTP}{File Transfer Protocol}
\newacronym{fw}{FW}{Flow Window}
\newacronym{ge}{GE}{Gaussian Elimination}
\newacronym{gnb}{gNB}{Next Generation Node Base}
\newacronym{gop}{GOP}{Group of Pictures}
\newacronym{gpr}{GPR}{Gaussian Process Regressor}
\newacronym{gpu}{GPU}{Graphics Processing Unit}
\newacronym{gtp}{GTP}{GPRS Tunneling Protocol}
\newacronym{gtpc}{GTP-C}{GPRS Tunnelling Protocol Control Plane}
\newacronym{gtpu}{GTP-U}{GPRS Tunnelling Protocol User Plane}
\newacronym{gtpv2c}{GTPv2-C}{\gls{gtp} v2 - Control}
\newacronym{gw}{GW}{Gateway}
\newacronym{harq}{HARQ}{Hybrid Automatic Repeat reQuest}
\newacronym{hetnet}{HetNet}{Heterogeneous Network}
\newacronym{hh}{HH}{Hard Handover}
\newacronym{hol}{HOL}{Head-of-Line}
\newacronym{hqf}{HQF}{Highest-quality-first}
\newacronym{hss}{HSS}{Home Subscription Server}
\newacronym{http}{HTTP}{HyperText Transfer Protocol}
\newacronym{hbf}{HBF}{Hybrid Beamforming}
\newacronym{ia}{IA}{Initial Access}
\newacronym{iab}{IAB}{Integrated Access and Backhaul}
\newacronym{ic}{IC}{Incident Command}
\newacronym{ietf}{IETF}{Internet Engineering Task Force}
\newacronym{imsi}{IMSI}{International Mobile Subscriber Identity}
\newacronym{imt}{IMT}{International Mobile Telecommunication}
\newacronym{iot}{IoT}{Internet of Things}
\newacronym{ip}{IP}{Internet Protocol}
\newacronym{itu}{ITU}{International Telecommunication Union}
\newacronym{kpi}{KPI}{Key Performance Indicator}
\newacronym{kpm}{KPM}{Key Performance Measurement}
\newacronym{kvm}{KVM}{Kernel-based Virtual Machine}
\newacronym{los}{LoS}{Line of Sight}
\newacronym{lsm}{LSM}{Link-to-System Mapping}
\newacronym{lstm}{LSTM}{Long Short Term Memory}
\newacronym{lte}{LTE}{Long Term Evolution}
\newacronym{lxc}{LXC}{Linux Container}
\newacronym{m2m}{M2M}{Machine to Machine}
\newacronym{mac}{MAC}{Medium Access Control}
\newacronym{manet}{MANET}{Mobile Ad Hoc Network}
\newacronym{mano}{MANO}{Management and Orchestration}
\newacronym{mc}{MC}{Multi-Connectivity}
\newacronym{mcc}{MCC}{Mobile Cloud Computing}
\newacronym{mchem}{MCHEM}{Massive Channel Emulator}
\newacronym{mcs}{MCS}{Modulation and Coding Scheme}
\newacronym{mec2}{MEC}{Multi-access Edge Computing}
\newacronym{mec}{MEC}{Mobile Edge Computing}
\newacronym{mfc}{MFC}{Mobile Fog Computing}
\newacronym{mgen}{MGEN}{Multi-Generator}
\newacronym{mi}{MI}{Mutual Information}
\newacronym{mib}{MIB}{Master Information Block}
\newacronym{miesm}{MIESM}{Mutual Information Based Effective SINR}
\newacronym{mimo}{MIMO}{Multiple Input, Multiple Output}
\newacronym{ml}{ML}{Machine Learning}
\newacronym{mlr}{MLR}{Maximum-local-rate}
\newacronym[plural=\gls{mme}s,firstplural=Mobility Management Entities (MMEs)]{mme}{MME}{Mobility Management Entity}
\newacronym{mmtc}{mMTC}{Massive Machine-Type Communications}
\newacronym{mmwave}{mmWave}{millimeter wave}
\newacronym{mpdccp}{MP-DCCP}{Multipath Datagram Congestion Control Protocol}
\newacronym{mptcp}{MPTCP}{Multipath TCP}
\newacronym{mr}{MR}{Maximum Rate}
\newacronym{mrdc}{MR-DC}{Multi \gls{rat} \gls{dc}}
\newacronym{mse}{MSE}{Mean Square Error}
\newacronym{mss}{MSS}{Maximum Segment Size}
\newacronym{mt}{MT}{Mobile Termination}
\newacronym{mtd}{MTD}{Machine-Type Device}
\newacronym{mtu}{MTU}{Maximum Transmission Unit}
\newacronym{mumimo}{MU-MIMO}{Multi-user \gls{mimo}}
\newacronym{mvno}{MVNO}{Mobile Virtual Network Operator}
\newacronym{nalu}{NALU}{Network Abstraction Layer Unit}
\newacronym{nas}{NAS}{Network Attached Storage}
\newacronym{nat}{NAT}{Network Address Translation}
\newacronym{nbiot}{NB-IoT}{Narrow Band IoT}
\newacronym{nfv}{NFV}{Network Function Virtualization}
\newacronym{nfvi}{NFVI}{Network Function Virtualization Infrastructure}
\newacronym{ni}{NI}{Network Interfaces}
\newacronym{nic}{NIC}{Network Interface Card}
\newacronym{now}{NOW}{Non Overlapping Window}
\newacronym{nsm}{NSM}{Network Service Mesh}
\newacronym{nr}{NR}{New Radio}
\newacronym{nrf}{NRF}{Network Repository Function}
\newacronym{nsa}{NSA}{Non Stand Alone}
\newacronym{nse}{NSE}{Network Slicing Engine}
\newacronym{nssf}{NSSF}{Network Slice Selection Function}
\newacronym{oai}{OAI}{OpenAirInterface}
\newacronym{oaicn}{OAI-CN}{\gls{oai} \acrlong{cn}}
\newacronym{oairan}{OAI-RAN}{\acrlong{oai} \acrlong{ran}}
\newacronym{oam}{OAM}{Operations, Administration and Maintenance}
\newacronym{ofdm}{OFDM}{Orthogonal Frequency Division Multiplexing}
\newacronym{olia}{OLIA}{Opportunistic Linked Increase Algorithm}
\newacronym{omec}{OMEC}{Open Mobile Evolved Core}
\newacronym{onap}{ONAP}{Open Network Automation Platform}
\newacronym{onf}{ONF}{Open Networking Foundation}
\newacronym{onos}{ONOS}{Open Networking Operating System}
\newacronym{oom}{OOM}{\gls{onap} Operations Manager}
\newacronym{opnfv}{OPNFV}{Open Platform for \gls{nfv}}
\newacronym{oran}{O-RAN}{Open RAN}
\newacronym{orbit}{ORBIT}{Open-Access Research Testbed for Next-Generation Wireless Networks}
\newacronym{os}{OS}{Operating System}
\newacronym{oss}{OSS}{Operations Support System}
\newacronym{pa}{PA}{Position-aware}
\newacronym{pase}{PASE}{Prioritization, Arbitration, and Self-adjusting Endpoints}
\newacronym{pawr}{PAWR}{Platforms for Advanced Wireless Research}
\newacronym{pbch}{PBCH}{Physical Broadcast Channel}
\newacronym{pcef}{PCEF}{Policy and Charging Enforcement Function}
\newacronym{pcfich}{PCFICH}{Physical Control Format Indicator Channel}
\newacronym{pcrf}{PCRF}{Policy and Charging Rules Function}
\newacronym{pdcch}{PDCCH}{Physical Downlink Control Channel}
\newacronym{pdcp}{PDCP}{Packet Data Convergence Protocol}
\newacronym{pdsch}{PDSCH}{Physical Downlink Shared Channel}
\newacronym{pdu}{PDU}{Packet Data Unit}
\newacronym{pf}{PF}{Proportional Fair}
\newacronym{pgw}{PGW}{Packet Gateway}
\newacronym{phich}{PHICH}{Physical Hybrid ARQ Indicator Channel}
\newacronym{phy}{PHY}{Physical}
\newacronym{pmch}{PMCH}{Physical Multicast Channel}
\newacronym{pmi}{PMI}{Precoding Matrix Indicators}
\newacronym{powder}{POWDER}{Platform for Open Wireless Data-driven Experimental Research}
\newacronym{ppo}{PPO}{Proximal Policy Optimization}
\newacronym{ppp}{PPP}{Poisson Point Process}
\newacronym{prach}{PRACH}{Physical Random Access Channel}
\newacronym{prb}{PRB}{Physical Resource Block}
\newacronym{psnr}{PSNR}{Peak Signal to Noise Ratio}
\newacronym{pss}{PSS}{Primary Synchronization Signal}
\newacronym{pucch}{PUCCH}{Physical Uplink Control Channel}
\newacronym{pusch}{PUSCH}{Physical Uplink Shared Channel}
\newacronym{qam}{QAM}{Quadrature Amplitude Modulation}
\newacronym{qci}{QCI}{\gls{qos} Class Identifier}
\newacronym{qoe}{QoE}{Quality of Experience}
\newacronym{qos}{QoS}{Quality of Service}
\newacronym{quic}{QUIC}{Quick UDP Internet Connections}
\newacronym{ra}{RA}{Resouces Allocation}
\newacronym{rach}{RACH}{Random Access Channel}
\newacronym{ran}{RAN}{Radio Access Network}
\newacronym[firstplural=Radio Access Technologies (RATs)]{rat}{RAT}{Radio Access Technology}
\newacronym{rbg}{RBG}{Resource Block Group}
\newacronym{rb}{RB}{Resource Block}
\newacronym{rcn}{RCN}{Research Coordination Network}
\newacronym{rc}{RC}{RAN Control}
\newacronym{rec}{REC}{Radio Edge Cloud}
\newacronym{red}{RED}{Random Early Detection}
\newacronym{renew}{RENEW}{Reconfigurable Eco-system for Next-generation End-to-end Wireless}
\newacronym{rf}{RF}{Radio Frequency}
\newacronym{rfc}{RFC}{Request for Comments}
\newacronym{rfr}{RFR}{Random Forest Regressor}
\newacronym{ric}{RIC}{\gls{ran} Intelligent Controller}
\newacronym{rlc}{RLC}{Radio Link Control}
\newacronym{rlf}{RLF}{Radio Link Failure}
\newacronym{rlnc}{RLNC}{Random Linear Network Coding}
\newacronym{rmr}{RMR}{RIC Message Router}
\newacronym{rmse}{RMSE}{Root Mean Squared Error}
\newacronym{rnis}{RNIS}{Radio Network Information Service}
\newacronym{rr}{RR}{Round Robin}
\newacronym{rrc}{RRC}{Radio Resource Control}
\newacronym{rrm}{RRM}{Radio Resource Management}
\newacronym{rru}{RRU}{Remote Radio Unit}
\newacronym{rs}{RS}{Remote Server}
\newacronym{rsrp}{RSRP}{Reference Signal Received Power}
\newacronym{rsrq}{RSRQ}{Reference Signal Received Quality}
\newacronym{rss}{RSS}{Received Signal Strength}
\newacronym{rssi}{RSSI}{Received Signal Strength Indicator}
\newacronym{rtt}{RTT}{Round Trip Time}
\newacronym{ru}{RU}{Radio Unit}
\newacronym{rus}{RSU}{Road Side Unit}
\newacronym{rw}{RW}{Receive Window}
\newacronym{rx}{RX}{Receiver}
\newacronym{s1ap}{S1AP}{S1 Application Protocol}
\newacronym{sa}{SA}{standalone}
\newacronym{sack}{SACK}{Selective Acknowledgment}
\newacronym{sap}{SAP}{Service Access Point}
\newacronym{sbt}{SBT}{Scheduler-based throttling}
\newacronym{sc2}{SC2}{Spectrum Collaboration Challenge}
\newacronym{scef}{SCEF}{Service Capability Exposure Function}
\newacronym{sch}{SCH}{Secondary Cell Handover}
\newacronym{scoot}{SCOOT}{Split Cycle Offset Optimization Technique}
\newacronym{sctp}{SCTP}{Stream Control Transmission Protocol}
\newacronym{sdap}{SDAP}{Service Data Adaptation Protocol}
\newacronym{sdk}{SDK}{Software Development Kit}
\newacronym{sdm}{SDM}{Space Division Multiplexing}
\newacronym{sdma}{SDMA}{Spatial Division Multiple Access}
\newacronym{sdn}{SDN}{Software-defined Networking}
\newacronym{sdr}{SDR}{Software-defined Radio}
\newacronym{seba}{SEBA}{SDN-Enabled Broadband Access}
\newacronym{sgsn}{SGSN}{Serving GPRS Support Node}
\newacronym{sgw}{SGW}{Service Gateway}
\newacronym{si}{SI}{Study Item}
\newacronym{sib}{SIB}{Secondary Information Block}
\newacronym{sinr}{SINR}{Signal to Interference plus Noise Ratio}
\newacronym{sip}{SIP}{Session Initiation Protocol}
\newacronym{siso}{SISO}{Single Input, Single Output}
\newacronym{sla}{SLA}{Service Level Agreement}
\newacronym{sm}{SM}{Service Model}
%\newacronym{smf}{SMF}{Session Management Function}
\newacronym{smo}{SMO}{Service Management and Orchestration}
%\newacronym{sms}{SMS}{Short Message Service}
\newacronym{smsgmsc}{SMS-GMSC}{\gls{sms}-Gateway}
\newacronym{snr}{SNR}{Signal-to-Noise-Ratio}
\newacronym{son}{SON}{Self-Organizing Network}
\newacronym{sptcp}{SPTCP}{Single Path TCP}
\newacronym{srb}{SRB}{Service Radio Bearer}
\newacronym{srn}{SRN}{Standard Radio Node}
\newacronym{srs}{SRS}{Sounding Reference Signal}
\newacronym{ss}{SS}{Synchronization Signal}
\newacronym{sss}{SSS}{Secondary Synchronization Signal}
\newacronym{st}{ST}{Spanning Tree}
\newacronym{svc}{SVC}{Scalable Video Coding}
\newacronym{tb}{TB}{Transport Block}
\newacronym{tcp}{TCP}{Transmission Control Protocol}
\newacronym{tdd}{TDD}{Time Division Duplexing}
\newacronym{tdm}{TDM}{Time Division Multiplexing}
\newacronym{tdma}{TDMA}{Time Division Multiple Access}
\newacronym{tfl}{TfL}{Transport for London}
\newacronym{tfrc}{TFRC}{TCP-Friendly Rate Control}
\newacronym{tft}{TFT}{Traffic Flow Template}
\newacronym{tgen}{TGEN}{Traffic Generator}
\newacronym{tip}{TIP}{Telecom Infra Project}
\newacronym{tm}{TM}{Transparent Mode}
\newacronym{to}{TO}{Telco Operator}
\newacronym{tr}{TR}{Technical Report}
\newacronym{trp}{TRP}{Transmitter Receiver Pair}
\newacronym{ts}{TS}{Technical Specification}
\newacronym{tti}{TTI}{Transmission Time Interval}
\newacronym{ttt}{TTT}{Time-to-Trigger}
\newacronym{tue}{TUE}{Test UE}
\newacronym{tx}{TX}{Transmitter}
\newacronym{uas}{UAS}{Unmanned Aerial System}
\newacronym{uav}{UAV}{Unmanned Aerial Vehicle}
\newacronym{udm}{UDM}{Unified Data Management}
\newacronym{udp}{UDP}{User Datagram Protocol}
\newacronym{udr}{UDR}{Unified Data Repository}
\newacronym{ue}{UE}{User Equipment}
\newacronym{uhd}{UHD}{\gls{usrp} Hardware Driver}
\newacronym{ul}{UL}{Uplink}
\newacronym{um}{UM}{Unacknowledged Mode}
\newacronym{uml}{UML}{Unified Modeling Language}
\newacronym{upa}{UPA}{Uniform Planar Array}
\newacronym{upf}{UPF}{User Plane Function}
\newacronym{urllc}{URLLC}{Ultra Reliable and Low Latency Communications}
\newacronym{usa}{U.S.}{United States}
\newacronym{usim}{USIM}{Universal Subscriber Identity Module}
\newacronym{usrp}{USRP}{Universal Software Radio Peripheral}
\newacronym{utc}{UTC}{Urban Traffic Control}
\newacronym{vim}{VIM}{Virtualization Infrastructure Manager}
\newacronym{vm}{VM}{Virtual Machine}
\newacronym{vnf}{VNF}{Virtual Network Function}
\newacronym{volte}{VoLTE}{Voice over \gls{lte}}
\newacronym{voltha}{VOLTHA}{Virtual OLT HArdware Abstraction}
\newacronym{vr}{VR}{Virtual Reality}
\newacronym{vran}{vRAN}{Virtualized \gls{ran}}
\newacronym{vss}{VSS}{Video Streaming Server}
\newacronym{v2x}{V2X}{vehicle-to-everything}
\newacronym{v2i}{V2I}{vehicle-to-infrastructure}
\newacronym{v2v}{V2V}{vehicle-to-vehicle}
\newacronym{v2n}{V2N}{vehicle-to-network}
\newacronym{wbf}{WBF}{Wired Bias Function}
\newacronym{wf}{WF}{Waterfilling}
\newacronym{wg}{WG}{Working Group}
\newacronym{wlan}{WLAN}{Wireless Local Area Network}
\newacronym{osm}{OSM}{Open Source \gls{nfv} Management and Orchestration}
\newacronym{pnf}{PNF}{Physical Network Function}
\newacronym{drl}{DRL}{Deep Reinforcement Learning}
\newacronym{mtc}{MTC}{Machine-type Communications}
\newacronym{osc}{OSC}{O-RAN Software Community}
\newacronym{mns}{MnS}{Management Services}
\newacronym{ves}{VES}{\gls{vnf} Event Stream}
\newacronym{ei}{EI}{Enrichment Information}
\newacronym{fh}{FH}{Fronthaul}
\newacronym{fft}{FFT}{Fast Fourier Transform}
\newacronym{laa}{LAA}{Licensed-Assisted Access}
\newacronym{plfs}{PLFS}{Physical Layer Frequency Signals}
\newacronym{ptp}{PTP}{Precision Time Protocol}
\newacronym{lidar}{LiDAR}{Light Detection And Ranging}
\newacronym{dem}{DEM}{Digital Elevation Model}
\newacronym{dtm}{DEM}{Digital Terrain Model}
\newacronym{dsm}{DEM}{Digital Surface Models}
\newacronym{ota}{OTA}{Over-The-Air}
% \newacronym{oai}{OAI}{OpenAirInterface}
\newacronym{ns}{NS}{Network Slicing}
\newacronym{ne}{NE}{Nash Equilibrium}
\newacronym{hf}{HF}{High Frequency}
\newacronym{noma}{NOMA}{Non-Orthogonal Multiple Access}
\newacronym{sre}{SRE}{Smart Radio Environment}
\newacronym{ris}{RIS}{Reconfigurable Intelligent Surface}
\newacronym{inp}{InP}{Infrastructure Provider}
\newacronym{smf}{SMF}{Slicing Magangement Framework}
\newacronym{nsn}{NSN}{Network Slicing Negotiation}
\newacronym{sms}{SMS}{Slicing MAC Scheduler}
\newacronym{brd}{BRD}{Best Response Dynamics}
\newacronym{dssbr}{DSSBR}{Double Step Smoothed Best Response}
\newacronym{poa}{PoA}{Price of Anarchy}
\newacronym{pos}{PoS}{Price of Stability}
\newacronym{milp}{MILP}{Mixed Integer-Linear Program}
\newacronym{pod}{PoD}{Price of DSSBR}
\newacronym{roc}{ROC}{Radio Overload Control}
\newacronym{ciot}{cIoT}{critical Internet of Things}
\newacronym{embbpr}{eMBB Pr.}{enhanced Mobile BroadBand Premium}
\newacronym{sps}{SPS}{Semi-persistent Scheduling}
\newacronym{cg}{CG}{Configured Grant}
\newacronym{embbbs}{eMBB Bs.}{enhanced Mobile BroadBand Basic}
\newacronym{en}{EN}{Edge Node}
\newacronym{ec}{EC}{Edge Computing}
\newacronym{sp}{SP}{Service Provider}
\newacronym{me}{ME}{Market Equilibrium}
\newacronym{so}{SO}{Social Optimum}
\newacronym{wso}{WSO}{Weighted Social Optimum}
\newacronym{ps}{PS}{Proportional Sharing}
\newacronym{eg}{EG}{Eisenberg-Gale program}
\newacronym{pe}{PE}{Pareto Efficiency}
\newacronym{nsw}{NSW}{Nash Social Welfare}
\newacronym{ef}{EF}{Envy-Freeness}
\newacronym{sub6}{sub6GHz}{Below 6GHz}
\newacronym{ncr}{NCR}{Network-Controlled Repeater}
\newacronym{nlos}{NLoS}{Non-Line of Sight}
\newacronym{src}{SRC}{Smart Radio Connection}
\newacronym{srd}{SRD}{Smart Radio Device}
\newacronym{cs}{CS}{Candidate Site}
\newacronym{tp}{TP}{Test Point}
\newacronym{fov}{FoV}{Field of View}
\newacronym{nrric}{near-RT RIC}{Near Real-time {RAN} Intelligent Controller}
\newacronym{e2ap}{E2AP}{E2 Application Protocol}
\newacronym{e2sm}{E2SM}{E2 Service Model}
\newacronym{nrtric}{non-RT RIC}{Non-Real-Time {RIC}}
\newacronym{itti}{ITTI}{Inter-task Interface}
\newacronym{bap}{BAP}{Backhaul Adaptation Protocol}
\newacronym{iabest}{IABEST}{Integrated Access and Backhaul Experimental large-Scale Tetbed}
\newacronym{teid}{TEID}{Tunnel Endpoint Identifier}
\newacronym{dlsch}{DL-SCH}{Downlink Shared Channel }
\newacronym{ulsch}{UL-SCH}{Uplink Shared Channel }
\newacronym{rsu}{RSU}{Road Side Unit}
\newacronym{its}{ITS}{Intelligent Transportation Systems}
\newacronym{vanet}{VANET}{Vehicular Ad-hoc Network}
\newacronym{dt}{DT}{Digital Twin}
\newacronym{ecc}{ECC}{Edge Computing Cluster}
\newacronym{o2i}{O2I}{Outdoor-to-indoor}
\newacronym{fwa}{FWA}{Fixed wireless access}
\newacronym{afc}{AFC}{Automated Frequency Coordinator}
\newacronym{bb}{BB}{baseband}
\newacronym{cpri}{CPRI}{Common Public Radio Interface}
\newacronym{ecpri}{eCPRI}{Enhanced Common Public Radio Interface}
\newacronym{re}{RE}{Resource Element}
\newacronym{hfcl}{HFCL}{High-Frequency Campus Lab}
\newacronym{au}{AU}{Antenna Unit}
\newacronym{5g-phy}{5G PHY}{5G Physical Layer}
\newacronym{fso}{FSO}{Free Space Optics}
\newacronym{eirp}{EIRP}{Effective Isotropic Radiated Power}
\title{Towards Smart Fronthauling Management: Experimental Insights from a 5G Testbed}  %Proposta da migliorare
%match capacity
%keep the pace
%wireless fronthaul quality vs RAN capacity
%RAN-assisted something
%Adaptive RAN capacity
% \raggedbottom

% author names and affiliations
% use a multiple column layout for up to three different
% affiliations
\author{\IEEEauthorblockN{Marcello Morini\textsuperscript{1}, Eugenio Moro\textsuperscript{1}, Ilario Filippini\textsuperscript{1}, Danilo De Donno\textsuperscript{2}, Salvatore Moscato\textsuperscript{2}, Antonio Capone\textsuperscript{1}}
\IEEEauthorblockA{\textsuperscript{1} DEIB, Politecnico di Milano, Milan, Italy - \textit{\{name.surname\}@polimi.it}}
\textsuperscript{2} Milan Research Center, Huawei Technologies Italia S.r.l., Milan, Italy - \textit{name.surname@huawei.com}\vspace{-1.5em}}
% \and
% for over three affiliations, or if they all won't fit within the width
% of the page, use this alternative format:
% 
%\author{\IEEEauthorblockN{Michael Shell\IEEEauthorrefmark{1},
%Homer Simpson\IEEEauthorrefmark{2},
%James Kirk\IEEEauthorrefmark{3}, 
%Montgomery Scott\IEEEauthorrefmark{3} and
%Eldon Tyrell\IEEEauthorrefmark{4}}
%\IEEEauthorblockA{\IEEEauthorrefmark{1}School of Electrical and Computer Engineering\\
%Georgia Institute of Technology,
%Atlanta, Georgia 30332--0250\\ Email: see http://www.michaelshell.org/contact.html}
%\IEEEauthorblockA{\IEEEauthorrefmark{2}Twentieth Century Fox, Springfield, USA\\
%Email: homer@thesimpsons.com}
%\IEEEauthorblockA{\IEEEauthorrefmark{3}Starfleet Academy, San Francisco, California 96678-2391\\
%Telephone: (800) 555--1212, Fax: (888) 555--1212}
%\IEEEauthorblockA{\IEEEauthorrefmark{4}Tyrell Inc., 123 Replicant Street, Los Angeles, California 90210--4321}}

% make the title area
\maketitle

\begin{abstract}
\color{black} The fronthaul connection is a key component of \gls{cran} architectures, consistently required to handle high capacity demands. However, this critical feature is at risk when the transport link relies on wireless technology. Fortunately, solutions exist to enhance the reliability of wireless links.
In this paper, we recall the theoretical fronthaul model, present a dynamic reconfiguration strategy and perform a conclusive experiment. %describe all the steps required to complete an experiment on 
Specifically, we showcase the setup of a wireless fronthaul testbed and discuss the resulting measurements. For this task, we leveraged the commercial hardware provided by the \gls{hfcl}, a private 5G network with \gls{mmwave} radio access interface. 
Our experiments provide original data on the fronthaul utilization in this real deployment, demonstrating both a good accordance with the theoretical model discussed in \cite{shapingWirelessFH} and the viability of one stabilizing solution.
\color{black}
\end{abstract}

\IEEEpeerreviewmaketitle

\vspace{-0.1cm}
\section{Introduction}
\glsresetall
\color{black}
The Centralized RAN paradigm is reshaping cellular networks with disruptive force. \color{black}
%The heightened emphasis on decentralization within operators, industry and the \gls{oran} Alliance highlights its essential role in the future of cellular networks. %\gls{ran} decentralization techniques streamline site engineering and provide revolutionary deployment flexibility opportunities.
%
The emphasis on decentralization, particularly within the framework of \gls{oran}, underscores its anticipated fundamental role in the future of cellular networks. Operators and the industry at large are recognizing the value of decentralization, driving advancements that offer significant benefits in network flexibility and efficiency.
\gls{ran} decentralization techniques streamline site engineering and provide revolutionary deployment flexibility opportunities. Modern Ethernet-based fronthaul interfaces, such as \gls{ecpri} and \gls{oran} Open Fronthaul, facilitate this flexibility, allowing operators to rethink their transport networks and introduce heterogeneity. This capability to reimagine transport networks is remarkably helpful in optimizing network performance and cost.

The newly introduced flexibility enables operators to explore transport network alternatives to dedicated fibers\cite{wfh_survey1}, including E-band radio links, extremely high-frequency point-to-multipoint radio links, and even free-space optics. These alternatives present a promising reduction in capital and operational expenditures, especially in areas where new fiber deployment is prohibitively expensive. The wireless-vs-wired technology comparison in \cite{wfh_cost_effectivenes} shows that, in appropriate conditions, wireless technologies can provide sufficient link capacity at a cost even lower than wired counterparts. However, despite these cost benefits, alternative wireless approaches often need to improve reliability compared to dedicated fiber connections. Due to adverse atmospheric conditions, high-frequency fronthaul links and \acrlong{fso} are susceptible to link quality degradation. Additionally, packet-switched fronthaul networks are vulnerable to congestion, further complicating their deployment.

\begin{comment}
Among all the pieces of \gls{cran}, the fronthaul link is a matter of attention, as modern Ethernet-based fronthaul interfaces (i.e., \gls{ecpri} and Open Fronthaul) facilitate flexibility and heterogeneity in transport technologies. \color{black}
This, in turn, lets operators explore transport link alternatives to dedicated fibers~\cite{wfh_survey1}, such as point-to-point or point-to-multipoint \gls{mmwave} wireless links and even free-space optics. These alternatives may significantly reduce costs, especially in areas where new fiber deployment is prohibitively expensive. 
The wireless-vs-wired technology comparison in \cite{wfh_cost_effectivenes} shows that, in appropriate conditions, wireless technologies can provide sufficient link capacity at a cost even lower than their wired counterparts. However, wireless approaches often suffers from more reliability issues compared to dedicated fiber connections. Due to adverse atmospheric conditions, high-frequency fronthaul links and \acrlong{fso} are susceptible to link quality degradation. \color{black} The same behaviour may recur when the fronthaul uses shared optical links, where congestion may occur. \color{black}
\end{comment}

Traditionally, the reliability issue in wireless fronthaul links has been mitigated by over-dimensioning, a strategy that is inefficient and costly, negating many of the positive aspects of alternative solutions.
Adaptive solutions present a more efficient and cost-effective approach. Previous work by the authors  has introduced a dynamic and adaptive mechanism that manages instantaneous access resources to accommodate transitory fronthaul capacity reductions caused by effects such as rain \cite{shapingWirelessFH}. \color{black}The potential demonstrated by the simulations justifies a comprehensive set of practical experiments, which is the focus of this paper.\color{black}

\begin{comment}
As RAN disaggregation continues to evolve, the distinction between access and fronthaul is expected to blur, exploiting the full potential of decentralization. However, the novelty of this topic presents a significant challenge: the need for experimental system-level end-to-end testbeds. These testbeds are essential for evaluating the practicality and effectiveness of proposed solutions in a realistic environment, complementing simulation-based analyses. 
\end{comment}

To perform measurement campaigns, authors leveraged \gls{hfcl}: a standard-compliant private 5G testbed, designed to support rigorous experiments. 
This network comprises a macro 5G \gls{gnb} operating at \gls*{mmwave}, covering Politecnico di Milano campus and the surrounding neighborhood. The base station is dedicated solely to research and experimentation, ensuring high experimental repeatability. The high capacity of \gls*{mmwave} let us test the fronthaul connection under extreme access loads. The deployed hardware and tailored setup facilitate testing across a broad range of controlled conditions. This setup positions \gls{hfcl} as one of the few facilities worldwide capable of precisely replicating typical adverse atmospheric effects on advanced \gls*{mmwave} 5G networks.

In this work, we revisit the theoretical foundations of fronthaul rate adaptation mechanisms and describe two experiments applying these principles in a real network environment.  Specifically, the first experiment demonstrates the feasibility of the \textit{Cell Reconfiguration} strategy, while the second validates the theoretical fronthaul rate model presented in \cite{shapingWirelessFH}.
This study aims to provide the first insights into the real-world performance of fronthaul systems in front of varying channel conditions that lead to capacity reduction, potentially contributing to more effective wireless \gls{cran} deployments.

The paper is structured as follows. The next section provides context on \gls{cran} fronthaul and section~\ref{sec:backgroundSplits} discussed the available literature on this theme. Section~\ref{sec:model} reviews theoretical fronthaul rate model and adaptation strategies. The experimental section follows, detailing the testbed setup in Section~\ref{sec:HFCL} and the experiments in Section~\ref{sec:numerical}. Finally, Section~\ref{sec:conclusions} concludes the paper.

\

\vspace{-0.5cm}
\section{Background on C-RAN and functional splits} \label{sec:backgroundSplits}
% - Intro al fronthaul
\begin{figure}
    \centering
    \includegraphics[width=0.97\linewidth]{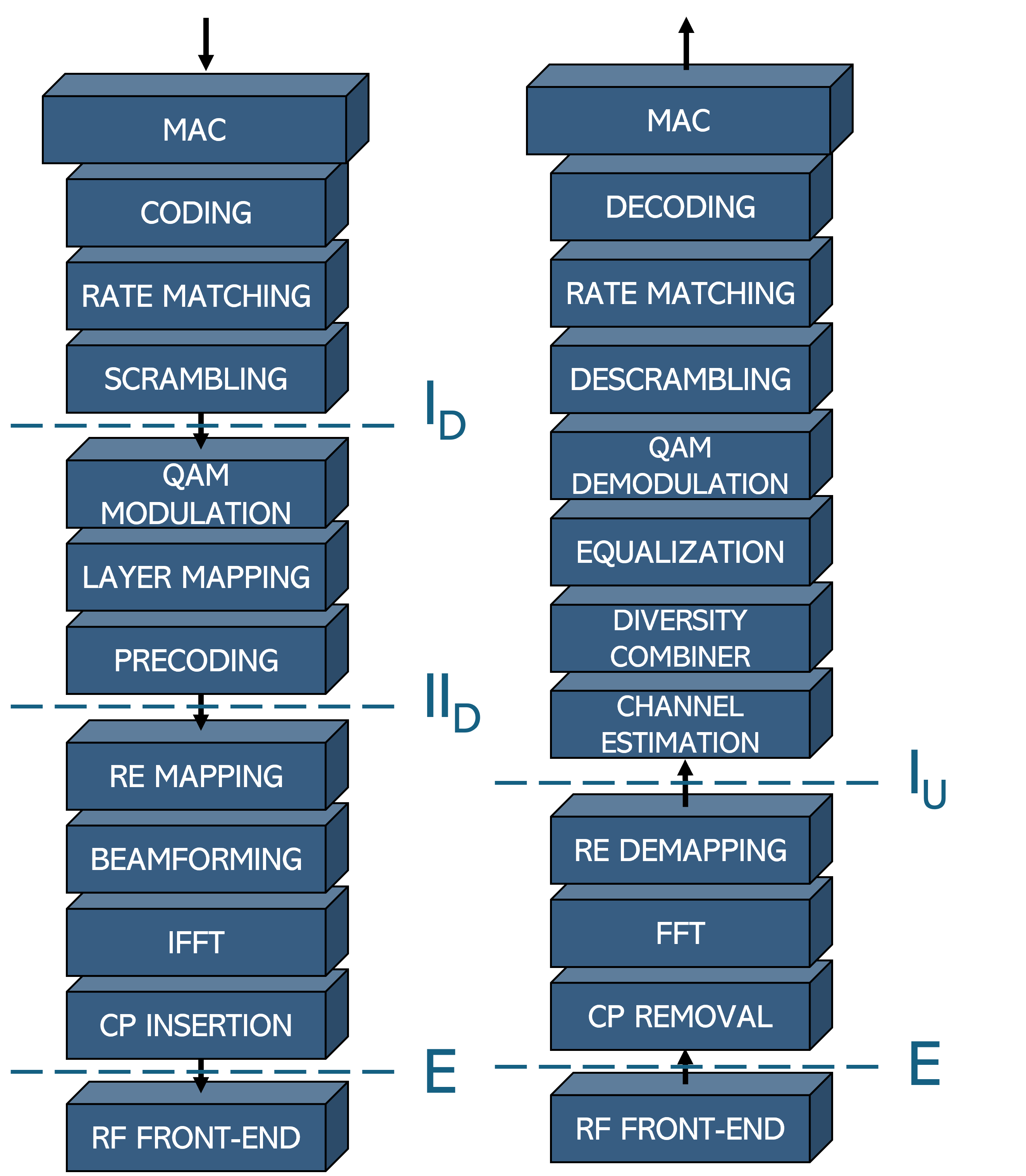}
    \caption{Physical layer functions chain and eCPRI splits \cite{eCPRI_standard}}
    \label{fig:phy_chain}
    \vspace{-0.5cm}
\end{figure}    
%TODO: If there's space, add description of the splits

\color{black}
The 5G \gls{gnb} implements an extensive chain of network functions. Part of this chain, namely the physical layer functions, are reported in Figure~\ref{fig:phy_chain}, for both \gls{ul} and \gls{dl} directions. Not all the functions are implemented in the same physical device, but they are split between two separate entities, connected via a fronthaul link. Depending on the reference architecture, these two entities have been named either \gls{aau} and \gls{bbu}, or \gls{ru} and \gls{du}. 
The decisio where to implement specific functions is known as \textit{split option}. 
Among the possible splits, the \gls{ecpri} standard defines five options -- three for \gls{dl} and two for \gls{ul} -- shown in Figure~\ref{fig:phy_chain}. For instance, if the network configuration chooses split option $I_D$, all functions from the RF front-end to the QAM modulation are implemented in the \gls{aau}, while higher-layer functions are handled in the \gls{bbu}. Recently, \gls{oran} Alliance has introduced additional split options, which expand upon the eCPRI-defined options \cite{oran_splits}.

Depending on the chosen split, different types of data are transmitted over the fronthaul. For example, basic bits are transmitted at split option $I_D$, while IQ symbols -- specifically bit-encoded IQ symbols -- are found at option $II_D$. Generally, the fronthaul data rate (or fronthaul rate) increases as the split moves from higher to lower layers in the chain, eventually saturating at \textit{split E}. At this level, all unused \gls{rb} are zero-padded, and the rate, reaching its peak, becomes constant. This behavior contrasts with splits above E, where the fronthaul rate depends on the data requested by users. This principle underpins the model presented in \cite{shapingWirelessFH} and partially reviewed in Section~\ref{sec:model}. While latency and coordination requirements also vary across split options, this paper wants to analyze the feasibility of adaptation techniques for wireless fronthaul, therefore these aspects fall outside the scope of this paper.
In this context, understanding how to model fronthaul link capacity -- and the extent to which real implementations align with these models -- is crucial.
%In the next section, other 
\color{black}

\section{Related works} \label{sec:related}
\color{black}
This paper investigates the dynamic reconfiguration of 5G cells to adapt to variable fronthaul capacity in wireless links. In this section, we address both the broader context of our research -- covering the theoretical foundations of \gls{cran} splits and \gls{ran} reconfiguration strategies -- and the availability of testbeds for the practical evaluation of the fronthaul aspects, which is our primary focus. 

% Theoretical Foundations of Fronthaul Capacity
Various studies have investigated the requirements of \gls{cran} splits.  For instance, \cite{survey_functional_split_2019} offers a comprehensive review of different functional splits, ranging from low-PHY to PDCP/RRC, highlighting their pros and cons along with formulas for estimating fronthaul rate requirements. Additionally, \cite{darts} presents simulation data to illustrate typical fronthaul rate needs.  Building on these foundational works, \cite{shapingWirelessFH} introduces a more general and updated model, providing new insights into fronthaul rate dimensioning. While these references offer critical theoretical context, they do not explore practical implementations.

% RAN Reconfiguration Concepts
Research on RAN reconfiguration underscores the potential to adjust \gls{cran} splits and resources to meet various requirements, such as power, latency, and fronthaul capacity. The concept of \textit{Joint Access and Fronthaul Coordination} has been examined as both a static optimization problem \cite{matoussi}, and a time-varying one \cite{alba, fh-aware_optimization, quantizationResolution}. While we acknowledge and share some foundational ideas from this research, the introduction of flexible splits complicates the reconfiguration process significantly. In contrast, our approach focuses specifically on modifying MIMO layers and resource blocks, offering a simpler strategy that can be readily implemented in commercial systems.

% \textcolor{red}{discussed in works like \cite{alba}, \cite{matoussi}, \cite{fh-aware_optimization}, \cite{fh-aware_bcn}, and \cite{quantizationResolution}}, addresses these challenges. Our approach is aligned with this research but focuses specifically on modifying MIMO layers and resource blocks based on fluctuating fronthaul capacity, as explained in \cite{shapingWirelessFH} and mentioned in the following section.

% Experimental Fronthaul Testbeds
Real-world testbeds remain underexplored. Notable contributions include \cite{testbed_mountaser} and \cite{testbed_salama}, which focus on fiber-based fronthaul deployments. The work in \cite{testbed_mountaser} presents an OAI-based testbed to evaluate fronthaul rate, latency, and jitter, while \cite{testbed_salama} measures data rate and processing power for various splits. The most recent study, \cite{challenges_opportunities_wfh}, combines modeling with experiments on a fixed-capacity wireless fronthaul link.

While these studies offer valuable insights, our work distinguishes itself by focusing on a fronthaul testbed constructed with commercial hardware. This emphasis on practical evaluation addresses a gap in the literature, enhancing the understanding of fronthaul performance in commercial 5G deployments through a comprehensive set of measurements.
%Additionally, while older studies like \cite{balaji}, \cite{nikaein}, and \cite{roldan} have contributed to the early understanding of fronthaul performance, they are now considered outdated in the context of the rapid advances in 5G technology.

In summary, while previous literature has established the theoretical and conceptual groundwork, this paper advances the state of the art by providing a comprehensive analysis of fronthaul performance on a testbed utilizing commercial equipment. This work helps bridge the gap between theory and practical application in CRAN architecture.
\color{black}

\section{Foundations of Radio Access Reconfiguration}    \label{sec:model}
To provide theoretical foundations to the management strategies that are the focus of the experiments, this section discusses a fronthaul rate model. This model derives from \cite{shapingWirelessFH} and only reports the formulas for \textit{split $I_D$} and $I_U$.

The rates that the fronthaul link must support are:
\begin{equation}
R_{FH}^{I_D} = N_{RB} \cdot N_{SC} \cdot N_{MIMO} \cdot Q_M \cdot\frac{1}{T_S}, \label{R_I_D}
\end{equation}
\begin{equation}
R_{FH}^{I_U} =  N_{RB} \cdot N_{SC} \cdot  N_{MIMO} \cdot N_{IQ} \cdot\frac{1}{T_S},    \label{R_I_U}
\end{equation}
for the \gls{dl} and the \gls{ul}, respectively. 
Both formulas consider:
$N_{RB}$, the number of used \glspl{rb} by the radio access interface,
$N_{SC}$, the number of subcarriers per \gls{rb} (i.e., twelve),
$N_{MIMO}$, the number of concurrently active \gls{mimo} layers, and
$T_S$, the duration of an OFDM symbol.
Two parameters differ between the two formulas above: $Q_M$, which represents the number of bits per modulated by each symbol, and $N_{IQ}$, the number of bits used to encode an IQ sample (i.e., the bitwidth). In other words, this means that the \gls{dl} fronthaul carries flows of raw data bits while the \gls{ul} fronthaul carries bit-encoded IQ symbols. Since $N_{IQ}$ is usually bigger than $Q_M$, the \gls{ul} fronthaul rate is usually higher than the \gls{dl} counterpart.

It is important to note that these formulas include an approximation. Beyond the \gls{re} data, the fronthaul link must also carry antenna control information, with beamforming control being particularly significant. Further details on this aspect are available in \cite{shapingWirelessFH}. %This is a constant bitrate that does not depend on the load generated by access traffic. In addition, current FR2 antenna systems have typically a hybrid architecture, which limits the amount of this control information. Therefore, we do not consider this contribution in the rest of the paper.
We do not consider this contribution further in the paper, as it is highly implementation-dependent. However, its impact is expected to be minor in our testbed, which uses hybrid beamforming and thereby significantly reduces the antenna control information compared to full-digital beamforming.

% As for the access, the formula for the overall data rate of the cell is obtained by subtracting from \ref{R_I_D} the amount of resources allocated to control channels (e.g., PBCH, PRACH) and coding. The formula below is derived from the one proposed by 3GPP \cite{TS38.306} and specializes for our system:
% \begin{equation}
% R_{ACC} = N_{RB} \cdot N_{SC} \cdot N_{MIMO} \cdot Q_M \cdot R_{MAX} \cdot f_{TDD} \cdot \frac{(1 - OH)}{T_S},  \label{R_ACC}   
% \end{equation}
% where $R_{MAX}$ is the coding rate, $OH$ is amount of resources assigned to the control channels and  $f_{TDD}$ explicits the portion of time devoted to \gls{ul} and to \gls{dl} in case of \gls{tdd} adoption.

% - Contromisure
\color{black}
The fronthaul link capacity must always meet or exceed the rates calculated for $N_{RB} = N^{MAX}_{RB}$ and $N_{MIMO} = N^{MAX}_{MIMO}$, representing the maximum number of active \glspl{rb} and \gls{mimo} layers in the radio access configuration. Ensuring this constraint is typically manageable when dedicated optical fiber links are used for transport; however, it becomes a significant technical challenge when links are wireless or fiber is shared, as link capacity is no longer under the designer’s control.

If the fronthaul capacity falls short of the rate requirements, the system may attempt to maintain an effective fronthaul connection by adjusting access resources, leveraging the relationship between access and fronthaul. Specific countermeasures can limit the number of \glspl{rb}, $N_{RB}$, and \gls{mimo} layers, $N_{MIMO}$, available to the gNB's scheduler \cite{shapingWirelessFH}.
In particular, the \textit{cell reconfiguration} can be implemented by selectively disabling certain radio access resources through software reconfiguration.
This impacts the total radio access capacity according to the formula \cite{TS38.306}:
\[
R_{ACC} = N_{RB} \cdot N_{SC} \cdot N_{MIMO} \cdot Q_M \cdot R_{MAX} \cdot f_{TDD} \cdot \frac{(1 - OH)}{T_S}
\]
where $R_{MAX}$ is the target code rate, $OH$, is the overhead due to the control channels (e.g., PBCH, PRACH), and $f_{TDD}$, is a factor dependent on the \gls{tdd} frame scheme (e.g., DDDSU) activated in the network.

An open issue remains in evaluating the impact and effectiveness of this strategy in a real network. In the next section, we present the experimental setup, and in Section \ref{sec:numerical}, we demonstrate both the viability of \textit{cell reconfiguration} and the consistency of the system behavior with Eq.~\ref{R_I_D}. 
\color{black}

\section{Experimental Setup}  \label{sec:HFCL}
\begin{figure*}
    \centering
    \includegraphics[width=0.91\linewidth]{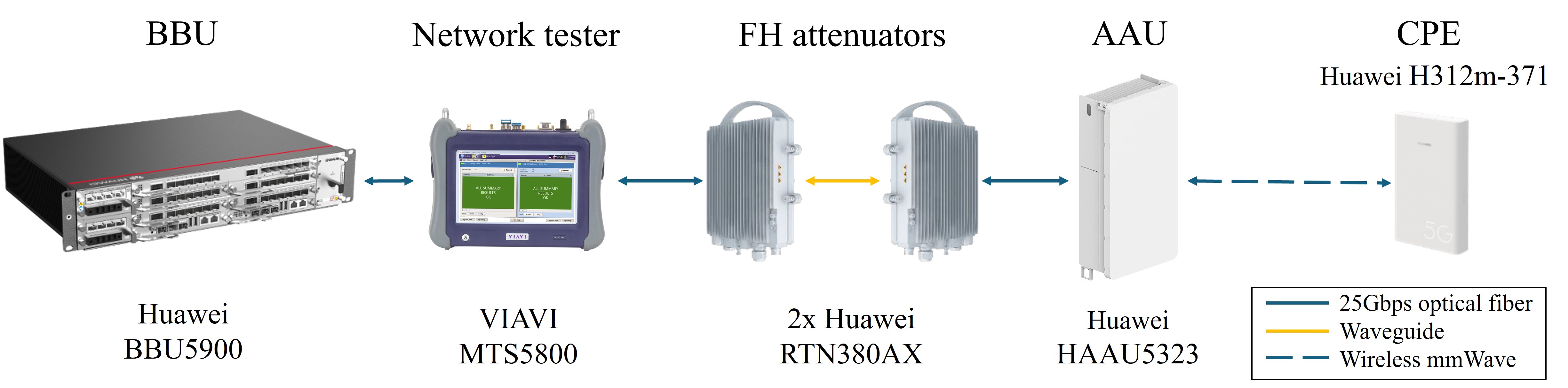}
    \caption{High-Frequency Campus Lab RAN equipment involved in the experiment}
    \label{fig:hfcl}
    % \vspace{-0.4cm}
\end{figure*} 

All the experiments were carried out at \gls{hfcl}, an open, experimental network deployed in the campus of Politecnico di Milano. It consists of a \acrlong{sa}, private 5G network equipped with tools for measuring and testing its performance. Its deployment is the result of the joint effort of Politecnico di Milano, telecommunication companies, equipment manufacturers, and service companies, to enhance cooperation among researchers.
The \gls{hfcl} radio access interface operates only at \acrshort{mmwave} and is only accessible from authorized \gls{cpe}. This allows us to produce high-fidelity 5G campaigns, excluding unwanted switching to other frequencies and interferences from external users.
%In the rest of this section, we illustrate the hardware hardware involved in the experiment.

\subsection{System architecture and hardware}
The full system used in the experiment is illustrated in Figure~\ref{fig:hfcl}.
The \gls{hfcl} includes an \acrfull{aau} \textit{Huawei HAAU5323}, connected to a centralized \acrfull{bbu} \textit{Huawei BBU5900}. The \gls{bbu} is in turn wired to a 5G virtualized core network (not reported in the picture). 
This equipment is commercially available and the underlying technology is proprietary. However, the possibility to access the \gls{gnb} configuration lets us modify many of its parameters, such as the access bandwidth, the \acrfull{tdd} structure and the transmitted power.
Details on its configuration are reported in \textit{Table \ref{tab:antenna_details}}. 
\begin{table}
\centering
\vspace{0.05in}
\begin{tabular}{|l|l|} 
\hline
\textbf{AAU coordinates (lat,lon,h)} & 45.478671, 9.232550, 22 m              \\ 
\hline
\textbf{Central AAU azimuth, down tilt} & 135°, 2°                               \\ 
\hline
\textbf{Center frequency}        & 27.2 GHz (n257)  \\ 
\hline
\textbf{Max channel bandwidth}       & 200 MHz          \\ 
\hline
\textbf{Subcarrier spacing}      & 120 kHz          \\ 
\hline
\textbf{TDD Frame structure}     & 4:1 (DDDSU) \\ 
\hline
\textbf{Maximum modulation (DL/UL)}     & 256 QAM/64 QAM                                \\ 
\hline
\textbf{Transmit power}            & 37.5 dBm         \\ 
\hline
\textbf{Gain}                & 32.5 dBi         \\ 
\hline
\textbf{\acrlong{eirp}}                & 70 dBm           \\ 
\hline
\textbf{Maximum MIMO layers}            & 8T-8R            \\ 
\hline
\textbf{Split (DL/UL)}                  & $I_D$/$I_U$       \\
\hline

\end{tabular}
\caption{\acrlong{aau} configuration parameters}
\label{tab:antenna_details}
\vspace{-0.2cm}
\end{table}

\begin{table}[]
\centering

\begin{tabular}{|l|l|}
\hline
\textbf{CPE model} & Huawei H312m-371  \\ \hline 
% \multirow{2}{*}{\textbf{Chipset}} & \multirow{2}{*}{\begin{tabular}[c|]{@{}l@{}}Huawei HiSilicon\\ Balong 5000 \end{tabular}} \\ \\ \hline
\textbf{Chipset}   & Huawei Hisilicon Balong 5000 \\ \hline
\textbf{Max EIRP {[}dBm{]}}& 40  \\ \hline
\textbf{MIMO layers }& 2T-2R  \\ \hline
\textbf{Antenna}& Planar array \\ \hline
\textbf{QAM order (DL/UL)} & 64/64  \\ \hline

\end{tabular}
\caption{\acrlong{cpe} characteristics}
\label{tab:CPE}
\vspace{-0.3cm}
\end{table}

The network is completed by a \gls{cpe} working as \gls{ue}, namely the \textit{Huawei H312m-371}. Its main characteristics are reported in \textit{Table \ref{tab:CPE}}.
To collect data transfer logs (e.g., received signal power), we connect the \glspl{cpe} to a PC running a \textit{drive test log tool}, namely the CPE-compatible \textit{Genex Probe 8.1} software. 
%Although many other details about \glspl{ue} can be given, we omit them for the sake of brevity as they are not relevant for analyses we performed through this paper's experiments, which mainly focus on the fronthaul link.

To carry out fronthaul experiments, the 5G network was complemented by additional devices. In order to emulate wireless links' capacity variations in our optical fiber fronthaul, we placed two \textit{Huawei RTN380AX} in the middle of the fronthaul optical connection. These are variable-rate transceivers working in the E-band (71–76 or 81–86 GHz) and usually employed in backhaul point-to-point connections. Among their features, they are capable to provide large capacities for wireless and support a wide set of channel bandwidths and modulations, that allow us adjusting E-band link capacity, symmetrically in both \gls{ul} and \gls{dl}, with a good granularity. In order to better confine and control the experiment, we connected the two transceivers with a waveguide instead of a free-space link. This optimizes space and allows high reproducibility of the experiments. \\
Finally, we included in the fronthaul chain a \textit{VIAVI MTS-5800} network tester. This device was inserted between the \gls{bbu} and the waveguide to measure the bitrate crossing the fronthaul link.
We leveraged this reconfiguration capabilities in the experiments of Section~\ref{sec:numerical}.
%https://carrier.huawei.com/en/products/wireless-network/microwave/e-band/rtn380ax
%https://www.viavisolutions.com/en-us/products/t-berd-mts-5800-platform#overview
%
% Summarizing, the radio access link is supported by \gls{mmwave} and the fronthaul link is ruled by an E-band guided radio transmission. 
%This is the equipment that let us carry out the experiment on the fronthaul link. In the next sections the experiment, the expected results and the measurements are discussed. %TODO bridge

\section{Experimental Results} \label{sec:numerical}
\color{black}
The experimental campaign consists of two experiments: the implementation of an example of \textit{cell reconfiguration} and a detailed \textit{analysis of the fronthaul rate requirements}.

\subsection{Implementation of cell reconfiguration}
\begin{table}[]
    \centering
    \begin{tabular}{|l|l|l|}
        \hline
        \textbf{Configurations}& \textbf{Channel bandwidth [MHz]} & \textbf{MIMO layers} \\ \hline 
         1& 200   & 4T-4R \\    \hline
         2& 100   & 4T-4R \\    \hline
         3& 100   & 8T-8R \\    \hline
    \end{tabular}
    \caption{Set of cell configurations}
    \label{tab:cell_config}
    \vspace{-0.6cm}
\end{table}
\begin{figure}
    \centering
    \includegraphics[width=1\linewidth]{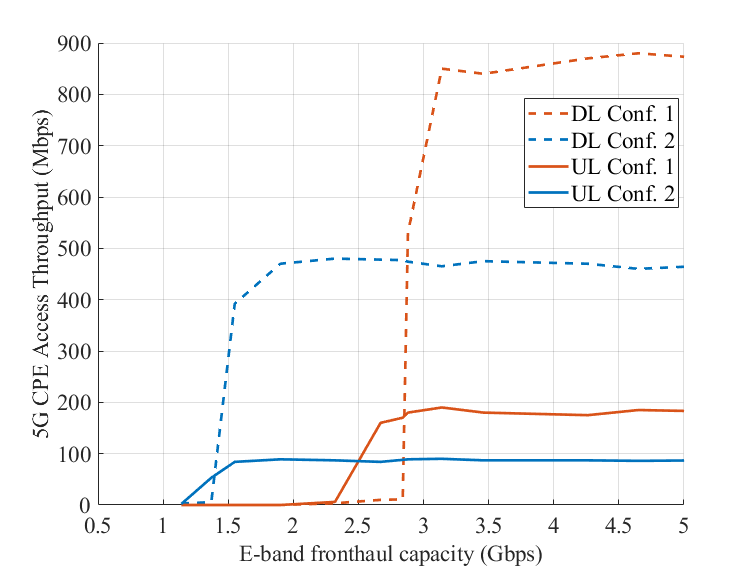}
    \caption{Impact of Cell Reconfiguration on fronthaul rates}
    \label{fig:fh_throttling}
    \vspace{-0.5cm}
\end{figure}
\begin{figure*}
\centering
  \begin{subfigure}[b]{0.49\textwidth}
    \includegraphics[width=\textwidth]{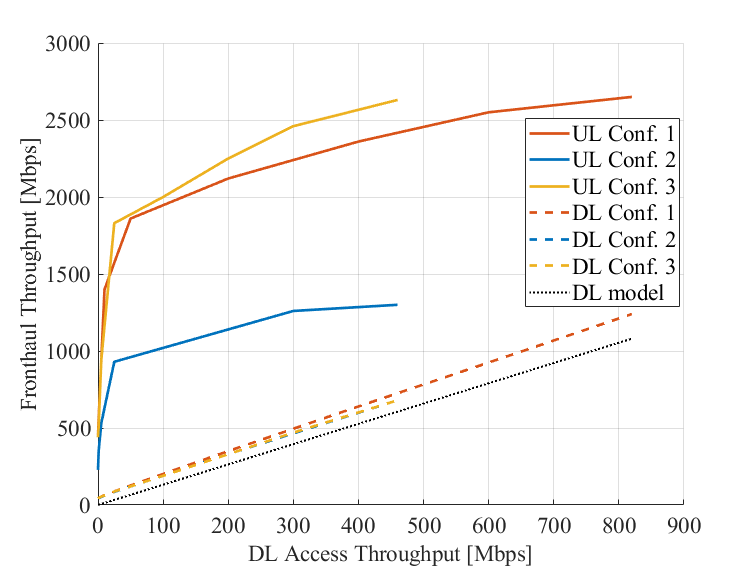}
    % \vspace{-10pt}
    \caption{}
    \label{fig:access_throttling_dl}  
  \end{subfigure}
 % \hfill
  \begin{subfigure}[b]{0.49\textwidth}
    \includegraphics[width=\textwidth]{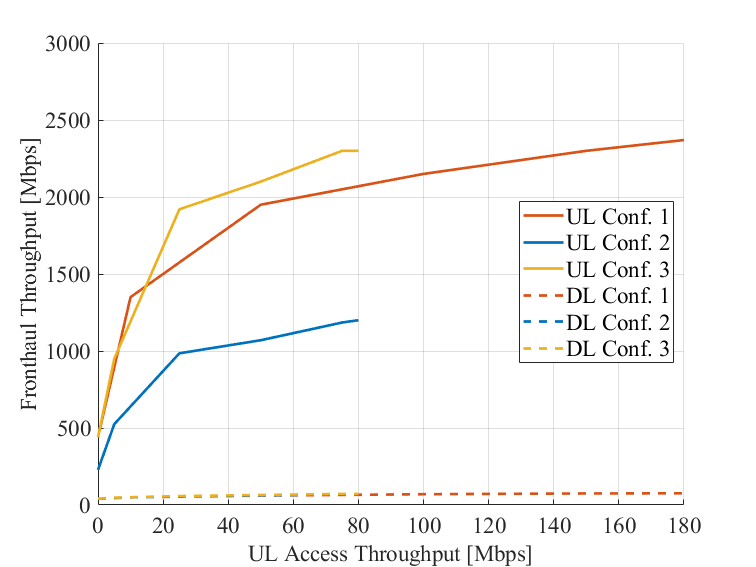}
    \vspace{-10pt}
    \caption{}
    \label{fig:access_throttling_ul}
  \end{subfigure}
\caption{Relation between fronthaul rate and access rate}
\label{fig:access_throttling}
\vspace{-0.3cm}
\end{figure*}
This experiment aims to demonstrate the feasibility of \textit{cell reconfiguration} adaptation strategy. To execute it, we leveraged two cell configurations: \textit{configuration 1} and \textit{2} reported in \textit{Table \ref{tab:cell_config}}.

The methodology used was the following. We enforced a bottleneck on the fronthaul capacity by setting different configurations of the E-band transceivers. By gradually increasing this capacity, we identified the minimum amount required to support the different cell configurations. We repeated this operation for each transmission direction and for every cell configuration. The results are shown in Figure~\ref{fig:fh_throttling}, where the achievable access throughput is plotted against the available fronthaul capacity. 

Configuration 1 shows that the UE can reach a \gls{dl} rate of 850~Mbps if the fronthaul supports at least 3.1~Gbps. Similarly, the uplink saturates at around 200~Mbps when the fronthaul is greater or equal than 2.7~Gbps.
The trends exhibit an on/off behavior: once a capacity threshold is reached, all available \glspl{rb} are allocated to the UE, allowing it to achieve maximum throughput.
In configuration 2, this pattern remains but is rescaled. he capacity threshold decreases to approximately 1.6~Gbps in both directions (i.e., about half of the previous case) at the cost of a halved access rate, which reaches 480~Mbps and 90~Mbps in \gls{dl} and \gls{ul} respectively. 
 Configuration 2 allocates only half of the \glspl{rb}, $N_{RB}$, available in configuration 1. This causes a reduction very close to a factor two not only in the in achievable access rate but also in \gls{dl} and \gls{ul} required fronthaul capacities, as predicted by the Eqs.~\ref{R_I_D} and \ref{R_I_U}. The slight deviation from the exact halved value results from additional rates generated for antenna control, which are implementation-dependent. Overall, the system behavior aligns with expectations, demonstrating the viability of the \textit{cell reconfiguration} strategy.
 
To further interpret the values in the plot, we conducted the following experiment.

\subsection{Analysis on the fronthaul rate requirements}
This experiment aims to deepen our understanding of fronthaul operations. Specifically, we show that for commonly considered fronthaul splits, the required fronthaul rate is closely linked to uplink traffic, with downlink traffic never acting as a bottleneck. For this analysis, we utilized all cell configurations listed in \textit{Table~\ref{tab:cell_config}}. 

We employed a methodology significantly different from the previous experiment. Without imposing any limitations on fronthaul capacity -- allowing the E-band link to operate at full capacity -- we used \textit{iperf3} to generate arbitrary bitrate values between two PCs: one connected to the CPE and the other to the N6 interface of the \gls{cn}. This setup allowed us to observe how the generated bitrate translates over the fronthaul.

These values are shown on the x- and y-axes in Figure~\ref{fig:access_throttling}. Note that in each plot, the traffic is actively generated only in one direction (i.e., \gls{dl} in \textit{Fig.~\ref{fig:access_throttling_dl}} and in \gls{ul} in \textit{Fig.~\ref{fig:access_throttling_ul}}) yet fronthaul traffic also appears in the opposite direction. This is due to acknowledgments and control signals, which play a fundamental role.

In Figure~\ref{fig:access_throttling_dl}, the lines at the bottom of the graph indicate a linear relationship between the \gls{dl} access throughput and the \gls{dl} fronthaul throughput. The rates for configurations 2 and 3 increase linearly from the origin, reaching a point where access rate is 460~Mbps and fronthaul rate is 680~Mbps, at which point the \gls{ue} utilizes all the \glspl{rb} it can support. Although configuration 3 supports up to 8 MIMO layers, the UE is hardware-limited to 2 layers, so all 132 \glspl{rb} are allocated to the CPE, with 66 per active MIMO layer \cite{TS38.104}. Therefore, in the \gls{dl}, configurations 2 and 3 perform similarly, each operating with a 100 MHz channel bandwidth.

Configuration 1, with a 200 MHz bandwidth, provides up to 132 \glspl{rb} per MIMO layer, doubling the \gls{dl} rates: \gls{dl} access rate reaches 820~Mbps, and \gls{dl} fronthaul rate reaches 1240~Mbps. These values closely align with those predicted by Equation \ref{R_I_D}, shown as the black line in the figure. The minor offset observed is attributed to antenna control data.

The \gls{ul} curves in Figure~\ref{fig:access_throttling_dl} exhibit a markedly different behavior from those in the \gls{dl}.
First, fronthaul rate values are significantly higher in the \gls{ul} than in the \gls{dl}, even though \gls{ul} access traffic mainly consists of acknowledgment traffic. This discrepancy arises from a lower \gls{ul} functional split compared to the \gls{dl}. Consequently, $N_{IQ}$-bit encoded symbols are transmitted over the \gls{ul} fronthaul link rather than basic data bits, as is the case for the \gls{dl}. It is difficult to estimate the exact value of $N_{IQ}$ as it depends on dynamic compression algorithms.

Second, the \gls{ul} exhibits two distinct regions: a steep rise at low access rates, driven by antenna control data, and a gradual increase at higher rates, where the impact of access resources becomes evident. This latter behavior reflects the bits carried in transmitted OFDM symbols across the whole bandwidth during a symbol period. Importantly, only symbols with no data across all subcarriers can be omitted from transmission over the fronthaul. Partially loaded OFDM symbols must be fully transmitted on the UL fronthaul link. This explains why, at low access rates, loaded symbols are primarily driven by signaling data, whereas at higher rates, the dependence on user data becomes more pronounced. Additionally, the slope of the curve is lower than in the \gls{dl} case, as new resources -- namely, additional symbols to be transferred through the fronthaul -- are only needed once the currently used symbols are fully utilized.

Analyzing the three \gls{ul} curves in Figure~\ref{fig:access_throttling_dl}, we observe that the values of configurations 2 and 3, which were equal in the \gls{dl}, now differ by a factor of two at the same access rates. At their maximum access \gls{dl} rate, configuration 2 achieves 1300~Mbps in the fronthaul, while configuration 3 reaches 2630~Mbps. TThis behavior is attributed to the number of available MIMO layers: used symbols are replicated across every antenna port and transmitted, irrespective of their activation in a specific layer. When comparing configuration 1 and configuration 3, instead, we find their fronthaul rates to be quite similar.  Configuration 1 has double the bandwidth but half the MIMO layers compared to configuration 3. This aligns with Eq.~\ref{R_I_U}: doubling $N_{RB}$ and halving $N_{MIMO}$ results in no change to the required fronthaul rate.

Overall, this indicates that transmitting \glspl{rb} in \gls{ul} necessitates sending a number of resources over the fronthaul link that is proportional to the product of total bandwidth and activated MIMO layers, accounting for allof \glspl{rb} present in the same symbol period. However, calculating the exact number of \glspl{rb} transmitted through the fronthaul is complex, as it depends on the scheduler's allocation of bits to symbols, influencing the resulting fully unloaded symbols that can be omitted.

%%%%%%%%%

In Figure~\ref{fig:access_throttling_ul}, an arbitrary \gls{ul} throughput is enforced in the access via \textit{iperf3} . The trends observed previously are extended here, with a significant contribution from the \gls{ul} traffic and a minor contribution from the resulting \gls{dl} traffic. The latter occupies a negligible share of the bandwidth, while the former remains the dominant contribution. Notably, the maximum UL fronthaul occupancy remains similar to that in Figure~~\ref{fig:access_throttling_dl}, despite the \gls{ul} access traffic being pushed to 180 Mbps in this scenario, compared to the acknowledgment traffic generated by an 850 Mbps \gls{dl} connection in the previous figure.
%highlighting a predominant dependence on the absolute value of the traffic, rather than the actual \gls{ul} access occupation.

One final observation clarifies the results obtained in the \textit{cell reconfiguration} experiment. Referring back to the thresholds in Figure~\ref{fig:fh_throttling}, we can now conclude that these thresholds are dictated by the \gls{ul} transmission, which requires approximately 2650 Mbps for an 850 Mbps DL access rate, rather than by the \gls{dl} transmission, which only necessitates 1240 Mbps.

The experiment reported in this section demonstrates, through real measurements, the feasibility of the \textit{cell reconfiguration} strategy and the overall correctness of Eqs.~\ref{R_I_D} and \ref{R_I_U} in describing the fronthaul resource requirements. While downlink fronthaul rates are easily predictable, \gls{ul} fronthaul rates are significantly influenced by the implementation of the eCPRI fronthaul interface and the scheduling policy governing access traffic.
Overall, our system demonstrates that fronthaul management for downlink traffic using \textit{split $I_D$} is highly efficient, while the \textit{split $I_U$} for uplink traffic is more resource-intensive, representing the primary limitation during a reduction in fronthaul capacity.
\color{black}

\section{Conclusions} \label{sec:conclusions}
In this paper, we explored the dynamic reconfiguration of 5G cells to adapt to variable fronthaul capacities in wireless links, focusing on the theoretical underpinnings of \gls{cran} splits and RAN reconfiguration strategies. Through comprehensive measurements conducted on a testbed built with commercial hardware, we provided valuable insights into fronthaul performance and its implications for real-world 5G deployments.

Our findings demonstrate that the fronthaul management for downlink traffic implementing \textit{split $I_D$} is highly efficient, while the uplink \textit{split $I_U$} proves to be more resource-intensive and serves as a critical limitation under reduced fronthaul capacity conditions. The experimental results validate the accuracy of our proposed models, particularly \textit{Eqs.}~\ref{R_I_D} and \ref{R_I_U}, in characterizing fronthaul resource requirements.

Moreover, we established that the required fronthaul rates are predominantly influenced by uplink traffic, highlighting the need for careful consideration of scheduling policies and the implementation of eCPRI interfaces. This work not only advances the theoretical understanding of fronthaul dynamics but also bridges the gap between theory and practical applications, providing a foundation for future research in adaptive strategies for 5G networks.

\section*{Acknowledgments}
\small{The research in this paper has been carried out in the framework of Huawei-Politecnico di Milano Joint Research Lab.}

\bibliographystyle{ieeetr}
\bibliography{Bibliography.bib}

% that's all folks
% \balance
\end{document}